\documentclass[aps,nofootinbib,floatfix,twocolumn,showpacs]{revtex4}
\usepackage{amsfonts}
\usepackage{amsmath}
\usepackage{amssymb}
\usepackage{graphicx}

\begin{document}

\title{Gravitational baryogenesis in Gauss-Bonnet braneworld cosmology}

\author{M. C. Bento}
\email{bento@sirius.ist.utl.pt} \affiliation{Departamento de F\'{\i}sica and
Centro de F\'{\i}sica Te\'orica de Part\'{\i}culas, Instituto Superior
T\'{e}cnico, Av. Rovisco Pais, 1049-001 Lisboa, Portugal}

\author{R. Gonz\'{a}lez Felipe}
\email{gonzalez@cftp.ist.utl.pt} \affiliation{Departamento de F\'{\i}sica and
Centro de F\'{\i}sica  Te\'orica de Part\'{\i}culas, Instituto Superior
T\'{e}cnico, Av. Rovisco Pais, 1049-001 Lisboa, Portugal}

\author{N. M. C. Santos}
\email{ncsantos@cftp.ist.utl.pt} \affiliation{Departamento de F\'{\i}sica and
Centro de F\'{\i}sica  Te\'orica de Part\'{\i}culas, Instituto Superior
T\'{e}cnico, Av. Rovisco Pais, 1049-001 Lisboa, Portugal}



\begin{abstract}
The mechanism of gravitational baryogenesis, based on the $CPT$-violating
gravitational interaction between the derivative of the Ricci scalar curvature
and the baryon-number current, is investigated in the context of the
Gauss-Bonnet braneworld cosmology. We study the constraints on the fundamental
five-dimensional gravity scale, the effective scale of $B$-violation and the
decoupling temperature, for the above mechanism to generate an acceptable
baryon asymmetry during the radiation-dominated era. The scenario of
gravitational leptogenesis, where the lepton-number violating interactions are
associated with the neutrino mass seesaw operator, is also considered.

\end{abstract}

\maketitle

\section{Introduction}

The origin of the baryon asymmetry is an outstanding problem in particle
physics and cosmology. Sufficient conditions for baryogenesis are the violation
of baryon number, the violation of $C$ and $CP$ symmetries and the existence of
nonequilibrium processes~\cite{Sakharov:dj}. Alternatively, if $CPT$ and baryon
number are violated, a baryon asymmetry could arise even in thermal
equilibrium~\cite{Dolgov:1981hv,Cohen:1987vi}.  In
Ref.~\cite{Bertolami:1996cq}, the effects on baryogenesis of certain
$CPT$-violating terms arising in a string-based framework were investigated and
it was shown that a large baryon asymmetry could be produced at the
grand-unified scale. Recently, a new baryogenesis mechanism, where the baryon
asymmetry is generated via a dynamical breaking of $CPT$ while maintaining
thermal equilibrium, was proposed in Ref.~\cite{Davoudiasl:2004gf}. The crucial
ingredient is an interaction between the derivative of the Ricci scalar
curvature ${\cal R}$ and the baryon-number ($B$) current $J^{\mu}$ (or any
current that leads to a net $B-L$ charge in equilibrium, where $L$ is the
lepton number, so that the asymmetry will not be erased by the electroweak
anomaly~\cite{Kuzmin:1985mm}):
\begin{align}
 \frac{1}{M_*^2} \int d^4 x \sqrt{-g}\, (\partial_{\mu} {\cal R})\,J^{\mu},
 \label{current}
\end{align}
where $M_*$ characterizes the scale of the interaction in the effective theory.
Such an operator is expected to arise in the low-energy effective field theory
of quantum gravity or in supergravity theories from a higher-dimensional
operator~\cite{Kugo:1982mr}.

The interaction in Eq.~(\ref{current}) violates $CP$ and, in an expanding
universe, it also dynamically breaks $CPT$. If one requires the existence of
$B$-violating processes in thermal equilibrium, then a net baryon asymmetry can
survive after their decoupling at a temperature $T_D$~\cite{Davoudiasl:2004gf},
\begin{align}
   \frac{n_B}{s} \propto \left.\frac{\dot{{\cal R}}}{M_*^2
    T}\right|_{T_D}\,. \label{asym}
\end{align}
For this mechanism to work, a nonvanishing time derivative $\dot{{\cal R}} \neq
0$ is necessary.\footnote{For a more general form of the derivative coupling of
the Ricci scalar to ordinary matter, ${\cal L} \propto \partial_\mu f({\cal R})
J^\mu$, see Ref.~\cite{Li:2004hh}.} Although in an expanding universe ${\cal R}
\sim H^2$ is nonzero in four-dimensional general relativity (GR), its time
derivative $\dot{{\cal R}}=0$ during the radiation-dominated (RD) epoch. It
turns out, however, that $\dot{{\cal R}} \neq 0$ can be easily realized in the
braneworld scenario, which suggests that higher-dimensional gravity effects can
offer a novel way to generate a baryon asymmetry through the dynamics of
spacetime~\cite{Shiromizu:2004cb}.

In the past few years, stimulated by the development of string theory, the
braneworld ideas, and particularly, the Randall-Sundrum (RS)
model~\cite{Randall:1999vf}, have been actively investigated. The RS braneworld
cosmology is based on the five-dimensional Einstein-Hilbert action; at high
energies, it is expected that this action will acquire quantum corrections, in
the form of higher-order curvature invariants in the bulk action. String theory
and holography indicate that such terms arise in the action at the perturbative
level. In five dimensions, the Gauss-Bonnet (GB) invariant has special
properties: it represents the unique combination that leads to second-order
gravitational field equations linear in the second derivatives and is
ghost-free~\cite{Nojiri:2000gv,Charmousis:2002rc,Lovelock:1971yv,Boulware:1985wk}.
Moreover, the graviton zero mode remains localized in the GB
braneworld~\cite{Mavromatos:2000az} and deviations from Newton's law at low
energies are less pronounced than in the RS case~\cite{Deruelle:2003tz}.

In this paper we examine gravitational baryogenesis in the context of GB
braneworld cosmology. The case when the GB contribution is absent and cosmology
is of RS type is also considered. We show that the observed baryon-to-entropy
ratio can be successfully explained in both frameworks. The possibility that
$B-L$ violation is associated with the neutrino mass seesaw operator is also
studied. In the latter case, the limits coming from low-energy neutrino physics
when combined with the GB inflationary constraints allow us to put bounds on
the fundamental scale of gravity, the effective scale of $B$-violation and the
decoupling temperature, which are required to generate an acceptable baryon
asymmetry in the gravitational leptogenesis scenario.

\section{Gauss-Bonnet braneworld}

The five-dimensional bulk action for the Gauss-Bonnet braneworld scenario is
given by
\begin{align}
{\cal S} &= \frac{1}{2\kappa_5^2} \int d^5x \sqrt{-\,^{(5)\!}g}
\left[-2\Lambda_5+ R \right. \nonumber\\
&\left.~{}+\alpha\, \left(R^2-4 R_{ab}R^{ab}+ R_{abcd}R^{abcd}\right) \right] \nonumber\\
&~{} - \int_{\rm brane} d^4x\, \sqrt{-g}\, \lambda + S_{\rm mat}\,,
\label{GBaction}
\end{align}
where $\alpha>0$ is the GB coupling, $\lambda>0$ is the brane tension,
$\Lambda_5<0$ is the bulk cosmological constant and $S_{\rm mat}$ denotes the
matter action. The fundamental energy scale of gravity is the 5D scale $M_5$
with $\kappa_5^2=8 \pi/ M_5^{3}$, and $M_4$ is the 4D Planck scale with
$\kappa_4^2=8 \pi/ M_4^{2}$.

The GB term may be viewed as the lowest-order stringy correction to the 5D
Einstein-Hilbert action with $\alpha \ll \ell^2$, where $\ell$ is the bulk
curvature scale, $|R|\sim 1/\ell^{2}$. The Randall-Sundrum type models are
recovered for $\alpha=0$. Moreover, for an anti-de Sitter bulk, it follows that
$\Lambda_5=-3\mu^2(2-\beta)$, where $\mu = 1/\ell$ is the extra-dimensional
energy scale and
\begin{align}
\beta \equiv 4\alpha\mu^2 \ll 1\,. \label{beta}
\end{align}

Imposing a $Z_2$ symmetry across the  brane in an anti-de Sitter bulk and
assuming that a perfect fluid matter source is confined to the brane, one gets
the modified Friedmann equation~\cite{Charmousis:2002rc,Lidsey:2003sj}
\begin{align}
\kappa_5^2(\rho+\lambda) = 2\mu\sqrt{1+\frac{H^2}{\mu^2}}\left[3-\beta +2 \beta
\frac{H^2}{\mu^2}\right]. \label{mf}
\end{align}
This can be rewritten in the useful form \cite{Lidsey:2003sj}
\begin{align}
H^2 = \frac{\mu^2}{\beta}\left[(1-\beta)\cosh\left(\!\frac{2\chi}{3}\!
\right)-1\right]\, , \label{hubble}
\end{align}
where  $\chi$ is a dimensionless measure of the energy density $\rho$
on the brane such that
\begin{align}
\rho+\lambda = m_\alpha^4 \sinh\chi\,,\label{chi}
\end{align}
with
\begin{align} \label{malpha}
m_\alpha=\left[\frac{8\mu^2(1-\beta)^3}{\beta \kappa_5^4}\right]^{1/8}
\end{align}
the characteristic GB energy scale. The GB high-energy regime ($\sinh\chi \gg
1$) corresponds then to $\rho \gg m_\alpha^4$. Notice also that we must have
$m_\alpha > m_\lambda=\lambda^{1/4}$, where $m_\lambda$ is the characteristic
RS energy scale. This in turn implies $\beta \lesssim
0.15$~\cite{Dufaux:2004qs}.

The requirement that one should recover general relativity at low energies
leads to the relation~\cite{Charmousis:2002rc,Maeda:2003vq}
\begin{align}
\kappa_4^2= \frac{\mu} {1+\beta}\, \kappa_5^2\,.\label{k4k5}
\end{align}
Since $\beta \ll 1$, we have $\mu \approx M_5^3/M_4^2$. Furthermore, the brane
tension is fine-tuned to achieve a zero cosmological constant on the
brane~\cite{Maeda:2003vq}:
\begin{align}
\kappa_5^2\lambda = 2\mu(3-\beta)\,.
\label{sig}
\end{align}

Expanding Eq.~(\ref{hubble}) in $\chi$, we find three regimes for the dynamical
history of the brane universe~\cite{Dufaux:2004qs}
\begin{align}
\rho\gg m_\alpha^4~& \Rightarrow ~ H^2\approx \left[
\frac{\mu^2\kappa_5^2}{4\beta}\, \rho \right]^{\!2/3} &\, ({\rm GB}),
\label{vhe}\\
m_\alpha^4 \gg \rho\gg m_\lambda^4~& \Rightarrow ~ H^2\approx
\frac{\kappa_4^2}{6\lambda}\, \rho^{2} &\, ({\rm RS}),\label{he}\\
\rho\ll m_\lambda^4~ & \Rightarrow ~ H^2 \approx \frac{\kappa_4^2}{3}\, \rho
&\, ({\rm GR}). \label{gr}
\end{align}
Eqs.~(\ref{vhe})-(\ref{gr}) are considerably simpler than the full Friedmann
equation and in many practical cases one of the three regimes can be assumed.
In this case, it is useful to consider a single patch with the effective
Friedmann equation~\cite{Calcagni:2004bh}
\begin{align}
H^2=\beta_q^2\, \rho^{q} \label{H2}\,,
\end{align}
where $q=1,~2,~2/3$ for GR, RS and GB regimes, respectively. For each regime,
the coefficients $\beta_q >0$ are determined in accordance with
Eqs.~(\ref{vhe})-(\ref{gr}).

\section{Gravitational baryogenesis}

Let us now consider the gravitational baryogenesis mechanism in the GB
braneworld. In an expanding universe, the interaction term in
Eq.~(\ref{current}) gives rise to an effective chemical potential $\mu_b \sim
\dot{{\cal R}}/M_*^2$ for baryons. In thermal equilibrium, the net
baryon-number density does not vanish as long as $\mu_b \neq 0$, and for $m_b,
\mu_b \ll T$ one has~\cite{Kolb:vq}
\begin{align}
n_B \approx \frac{g_b}{6} \mu_b T^2~,
\end{align}
where $g_b$ is the number of intrinsic degrees of freedom of the baryon. During
the RD epoch, this leads to a baryon-to-entropy ratio given by
\begin{align}
\frac{n_B}{s}=\left. -c \frac{\dot{\cal R}}{M_*^2 T}\right|_{T_D},\label{YB}
\end{align}
with
\begin{align}
c = \frac{15\, g_b}{4\pi^2 g_{\ast s}}~.
\end{align}
We have used $s=2 \pi^2 g_{\ast s} T^3/45$ for the entropy density, where
$g_{\ast s}$ is the total number of degrees of freedom which contribute to the
entropy of the universe.

The Ricci scalar in the Friedman-Robertson-Walker brane is defined in terms of
the expansion law as
\begin{align}
{\cal R} = -6\, ({\dot H}+2 H^2)~.
\end{align}
Using the full GB Friedmann Eq.~(\ref{hubble}), we get for the time derivative
of the Ricci scalar,
\begin{align}\label{Rdotfull}
   {\dot{\cal R}} &= -\frac{4\mu^2 (1+w)(1-\beta)}{\beta\,m_\alpha^4}\frac{H
   \rho}{\cosh\chi} \left[r_1(\chi)+r_2(\chi)\right]\,,\nonumber\\
   r_1(\chi) &= \left[-6+\frac{9}{2}(1+w)\right]
   \sinh\left(\!\frac{2\chi}{3}\!\right)\,,\nonumber\\
  r_2(\chi) & = 3 (1+w)\frac{\rho}{m_\alpha^4\,\cosh\chi}\nonumber\\
  & \times \left[\cosh\left(\!\frac{2\chi}{3}\!\right)
-\frac{3}{2}\sinh\left(\!\frac{2\chi}{3}\!\right)\tanh\chi\right]~,
\end{align}
where $w=p/\rho$ is the equation of state. During the RD era, $w=1/3$,
$r_1(\chi) =0$ and Eq.~(\ref{Rdotfull}) simplifies to
\begin{align}
{\dot{\cal R}}&= -\frac{32 \mu^2 (1-\beta)}{\beta~ m_\alpha^8}
\frac{H \rho^2}{\cosh^2\chi}  \nonumber\\
& \times \left[\frac{2}{3}\cosh\left(\!\frac{2\chi}{3}\!\right)
-\sinh\left(\!\frac{2\chi}{3}\!\right)\tanh\chi\right]~. \label{dotr}
\end{align}

\begin{figure*}[htb!]
\begin{center}
\includegraphics[height=6cm,width=7.5cm]{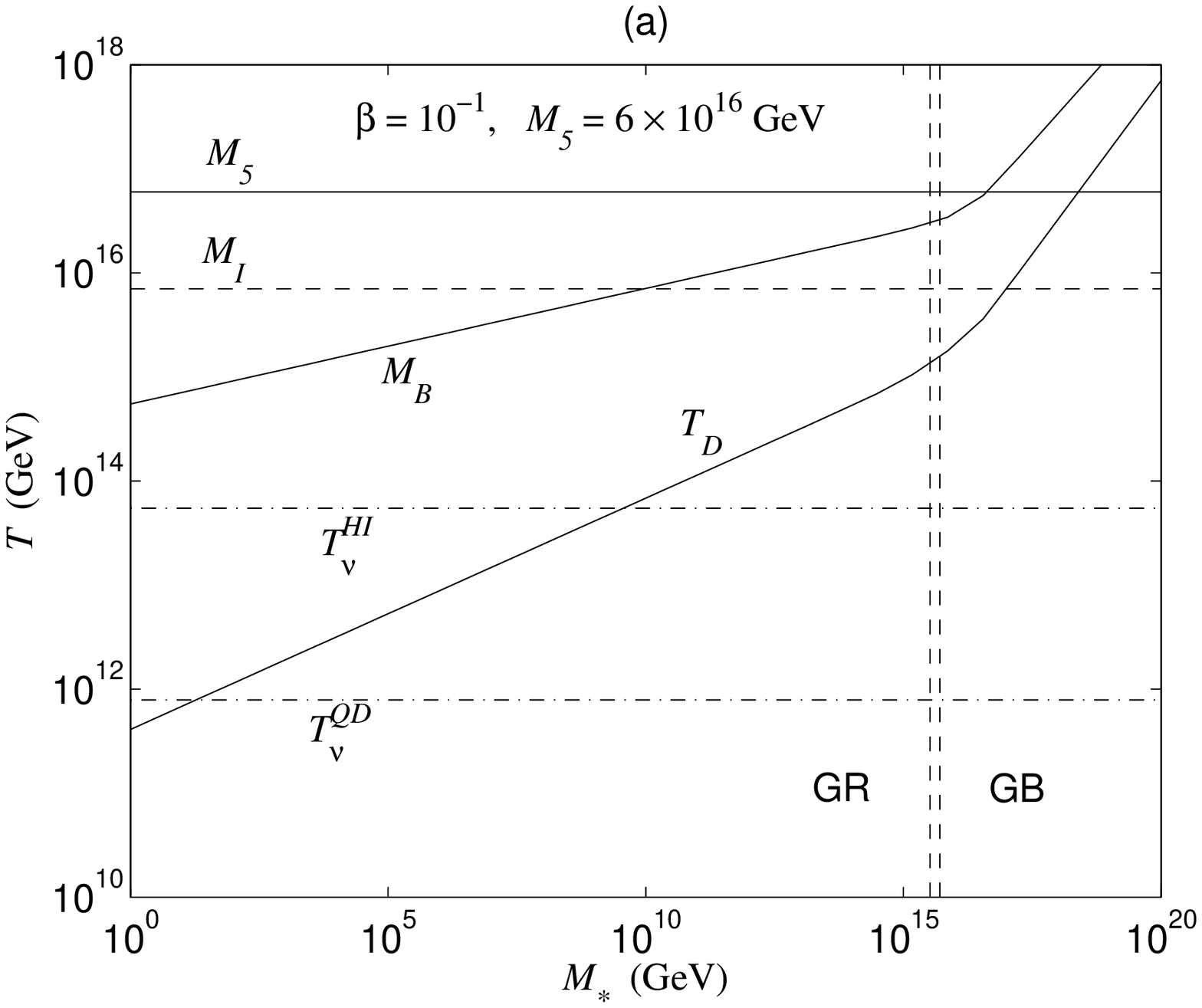}
\includegraphics[height=6cm,width=7.5cm]{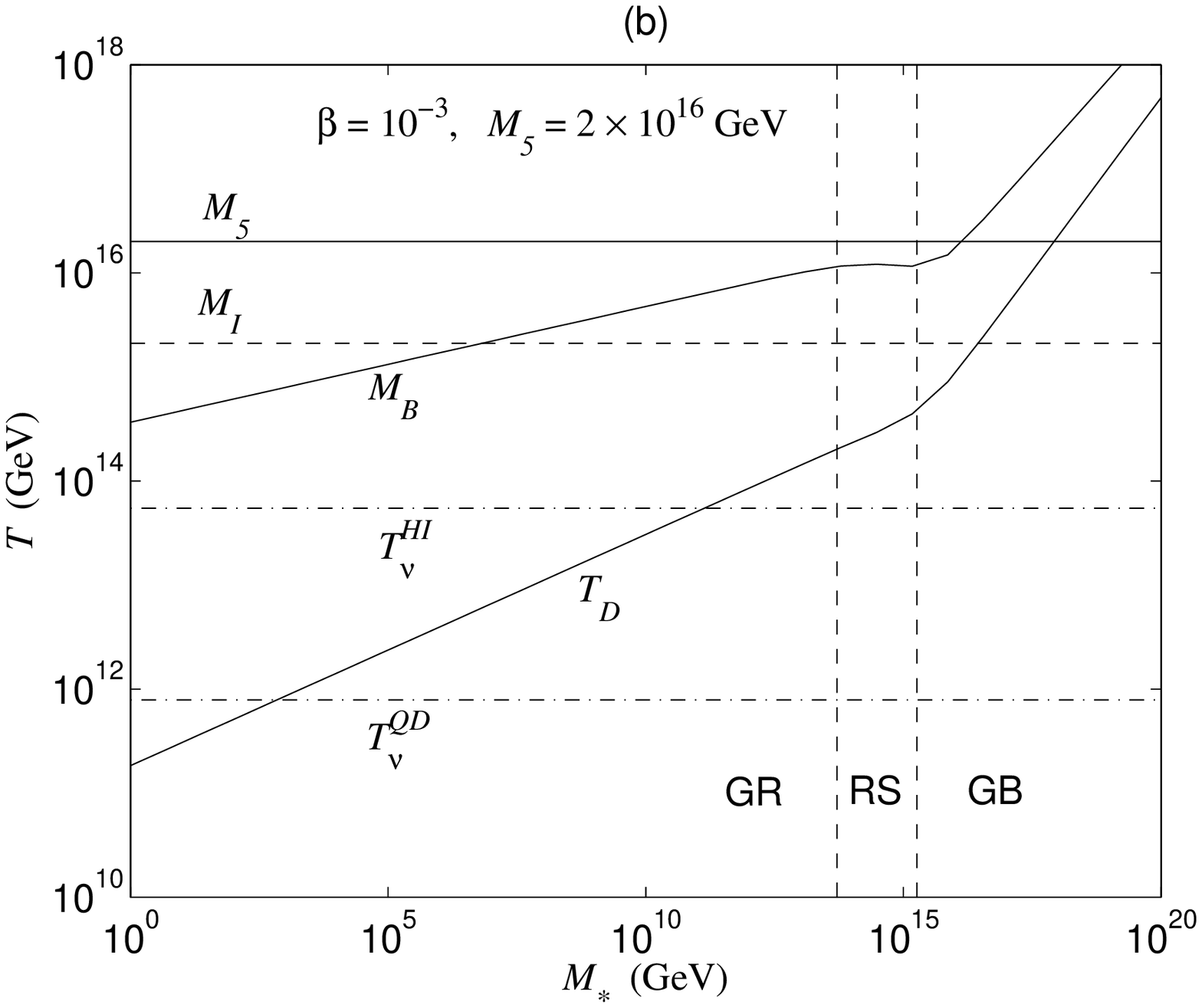}
\includegraphics[height=6cm,width=7.5cm]{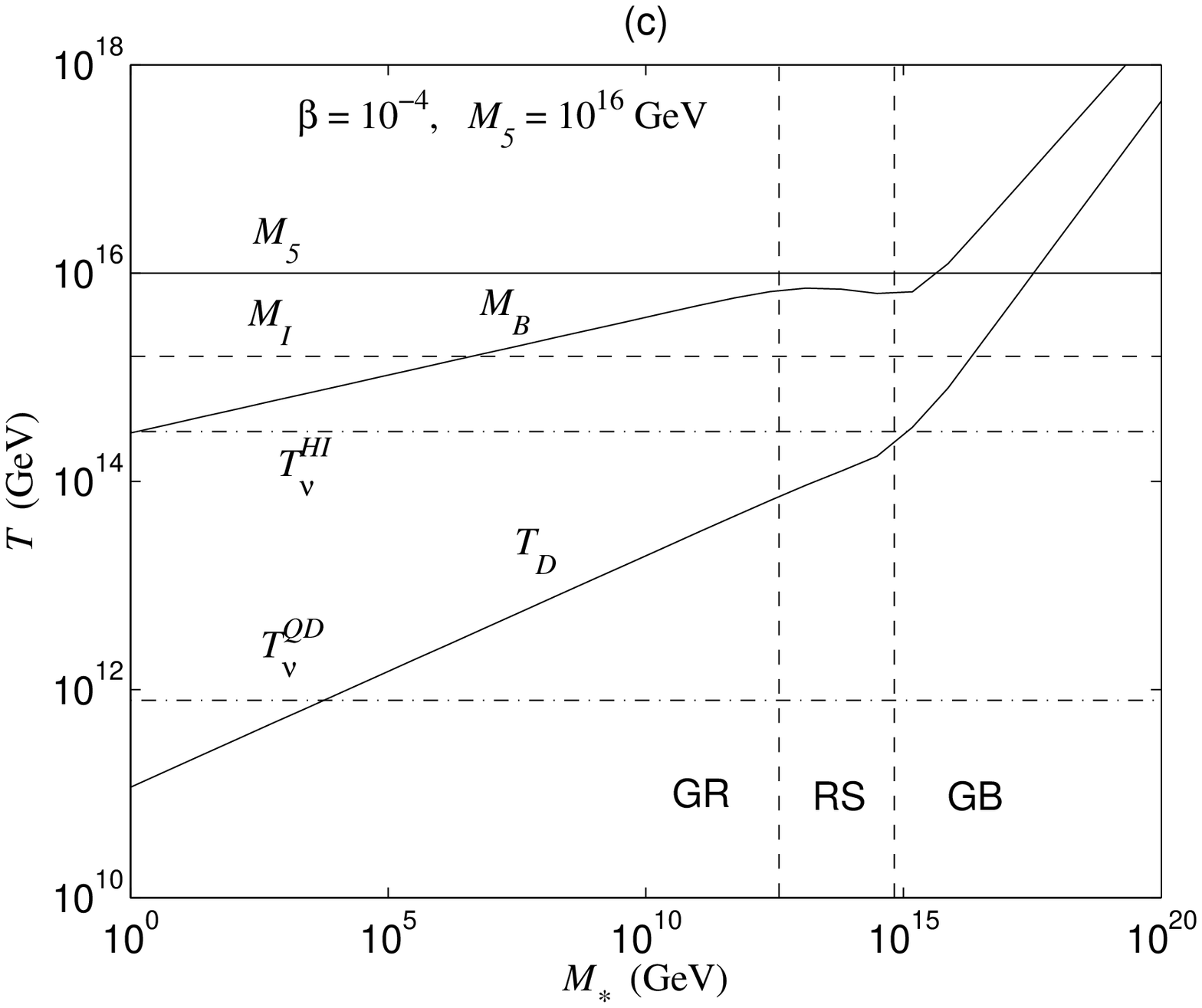}
\includegraphics[height=6cm,width=7.5cm]{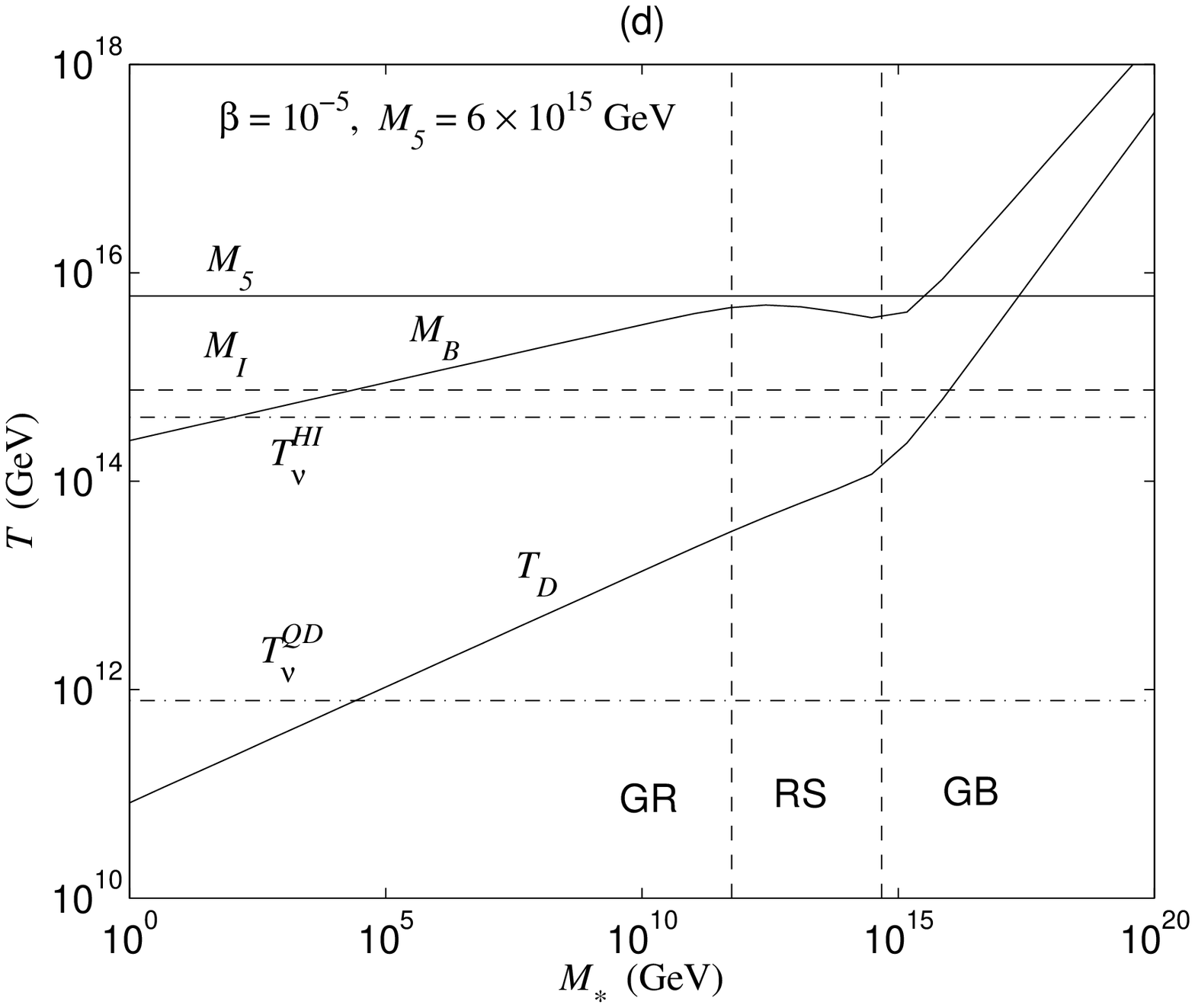}
\caption{\label{fig1} The decoupling temperature $T_D$ and the scale $M_B$ of
$B-L$ violation as functions of $M_\ast$ for different values of the
Gauss-Bonnet coupling $\beta$ and the fundamental scale of gravity $M_5$. The
vertical dashed lines delimit the transitions between the GB, RS and GR
regimes. The inflation scale $M_I$ (dashed line) and the decoupling
temperatures (dot-dashed lines) associated with the neutrino mass seesaw
operator in the case of quasidegenerate ($T_\nu^{QD}$) and hierarchical
($T_\nu^{HI}$) neutrino masses are also shown.}
\end{center}
\end{figure*}

Combining Eqs.~(\ref{YB}) and (\ref{dotr}), it is then possible to compute the
decoupling temperature, $T_D$, required to produce an acceptable
baryon-to-entropy ratio $n_B/s$. In Fig.~\ref{fig1} we plot this temperature as
a function of the scale $M_*$ for different values of the GB coupling $\beta$
and the fundamental gravity scale $M_5$, assuming the observed value $n_B/s
\simeq 9\times 10^{-11}$~\cite{Spergel:2003cb}.  We have used the fact that in
the RD era the energy density is $\rho=\pi^2 g_* T^4/ 30 $, where $g_*$ is the
total number of relativistic degrees of freedom. In the standard model $g_\ast
\simeq g_{\ast s} \simeq 100$ above the electroweak scale and, assuming $g_b
\sim {\cal O}(1)$, one gets $c \sim {\cal O}(10^{-2})$.

In order to establish the transition between the different regimes, we can
consider the simplified expansion law given by Eq.~(\ref{H2}). In this case,
\begin{align}
\dot{{\cal R}}=-48\, q\, (q-1)\, H^3 \label{Rdot}
\end{align}
and the baryon asymmetry reads as
\begin{align}
\frac{n_B}{s}= 48\,c\, q\, (q-1)\left(\!\frac{\pi^{2} g_*}{30}\!\right)^{3
q/2}\! \beta_q^3 \frac{T_D^{6 q-1}}{M_*^2}\,. \label{bara1}
\end{align}

If the decoupling occurs in the RS regime, where $q=2$ and
$\beta_q=(\kappa_4^2/6\lambda)^{1/2}\,$, we obtain
\begin{align}
T_D \simeq 3.2 \times 10^{-2}\left(M_\ast^2 \,M_5^9 \right)^{1/11}, \label{TD}
\end{align}
for $\beta \ll 1$. For the decoupling to occur in this regime, it is required
that $\rho(T_D)\geq m_\lambda^4\,$, which implies in turn that $M_* \geq
M_*^{RS}$ with
\begin{align}
M_\ast^{RS} \simeq 1.9 \times 10^5\left(\!\frac{M_5}{M_4}\!\right)^{\!11/4}
\!\! M_5\,. \label{mstarminRS}
\end{align}
On the other hand, in the GB regime, where $q=2/3$ and
$\beta_q=\left(\mu^2\kappa_5^2/4\beta\right)^{1/3}$, the decoupling temperature
is given by
\begin{align}
T_D \simeq 1.6 \times 10^{-4} \frac{\left( \beta M_\ast^2
M_4^4\right)^{1/3}}{M_5}\,, \label{TDrs}
\end{align}
so that the transition from the GB to the RS regime, defined by the condition
$\rho(T_D)=m_{\alpha}^4\,$, occurs for
\begin{align}
M_\ast^{GB}\simeq \frac{5.9 \times 10^4}{\beta^{11/16}}
\left(\!\frac{M_5}{M_4}\!\right)^{\!11/4} \!\! M_5\,. \label{mstarminGB}
\end{align}
The transition values $M_\ast^{RS}$ and $M_\ast^{GB}$ are represented in
Fig.~\ref{fig1} by the two vertical dashed lines. We see that the decoupling of
the baryon-number violating interactions can generally occur in either (GB, RS
or GR) regime. We also notice that the RS transition region shrinks as the
Gauss-Bonnet coupling $\beta$ increases.

An important constraint on the decoupling temperature comes from reheating and
inflation. Obviously, we must require $T_D < T_{rh} < M_I$, where $T_{rh}$ is
the reheating temperature (at which the universe becomes radiation dominated)
and $M_I$ is the inflation scale. In the conventional scenario, reheating
occurs as the inflaton field oscillates around its minimum and decays into
matter. In this case, $T_{rh}$ crucially depends on the details of the inflaton
coupling to matter. Here we restrict our discussion to the inflation scale
$M_I$. Assuming that inflation occurs in the high-energy (RS or GB) regime of
the theory and that it is driven by a quadratic potential, we find $M_I \approx
m_\alpha \sinh^{1/4} \!\chi_e\,$, where $\chi_e$ is evaluated at the end of
inflation (see the Appendix for details). In Fig.~\ref{fig1}, we have plotted
$M_I$ for different values of $\beta$ and $M_5$. We note that for $M_* \leq
M_5$ the constraint $T_D < M_I$ is always verified.

\begin{figure*}[htb!]
\begin{center}
\includegraphics[width=16cm]{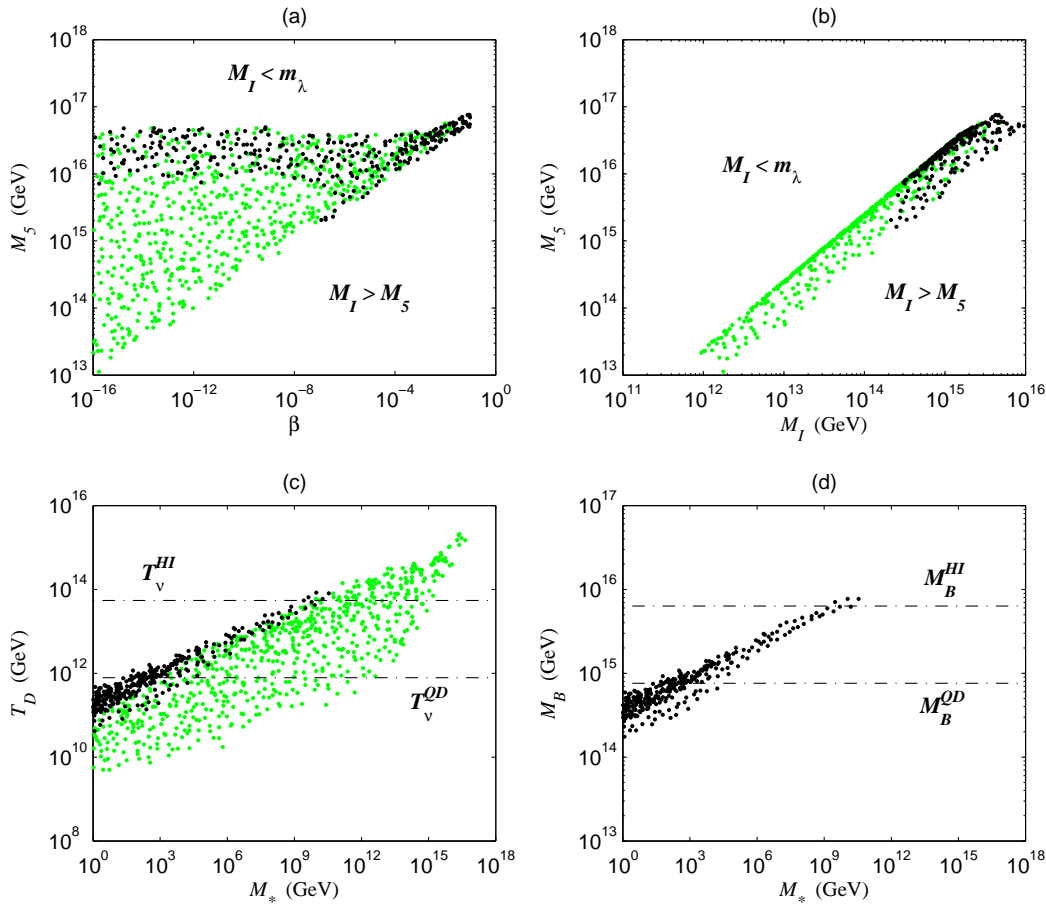}
\caption{\label{fig2} The parameter space that generates an observationally
acceptable baryon asymmetry during the radiation era in Gauss-Bonnet braneworld
cosmology with $T_D<M_I<M_5$. We take $n_B/s = 9 \times 10^{-11}$ and assume
that inflation is driven by a quadratic potential in the high-energy (RS or GB)
regime of the theory. The black-dotted region is obtained by taking into
account the constraint $M_B < M_I$ for the mass scale associated with a
dimension five $B$-violating operator. The horizontal dot-dashed lines
correspond to the GR decoupling temperatures $T_\nu^{QD}$ and $T_\nu^{HI}$
(panel (c)) and the scales $M_B^{QD}$ and $M_B^{HI}$ (panel (d)) associated
with the neutrino mass seesaw operator.}
\end{center}
\end{figure*}

\begin{figure*}[htb!]
\begin{center}
\includegraphics[width=16cm]{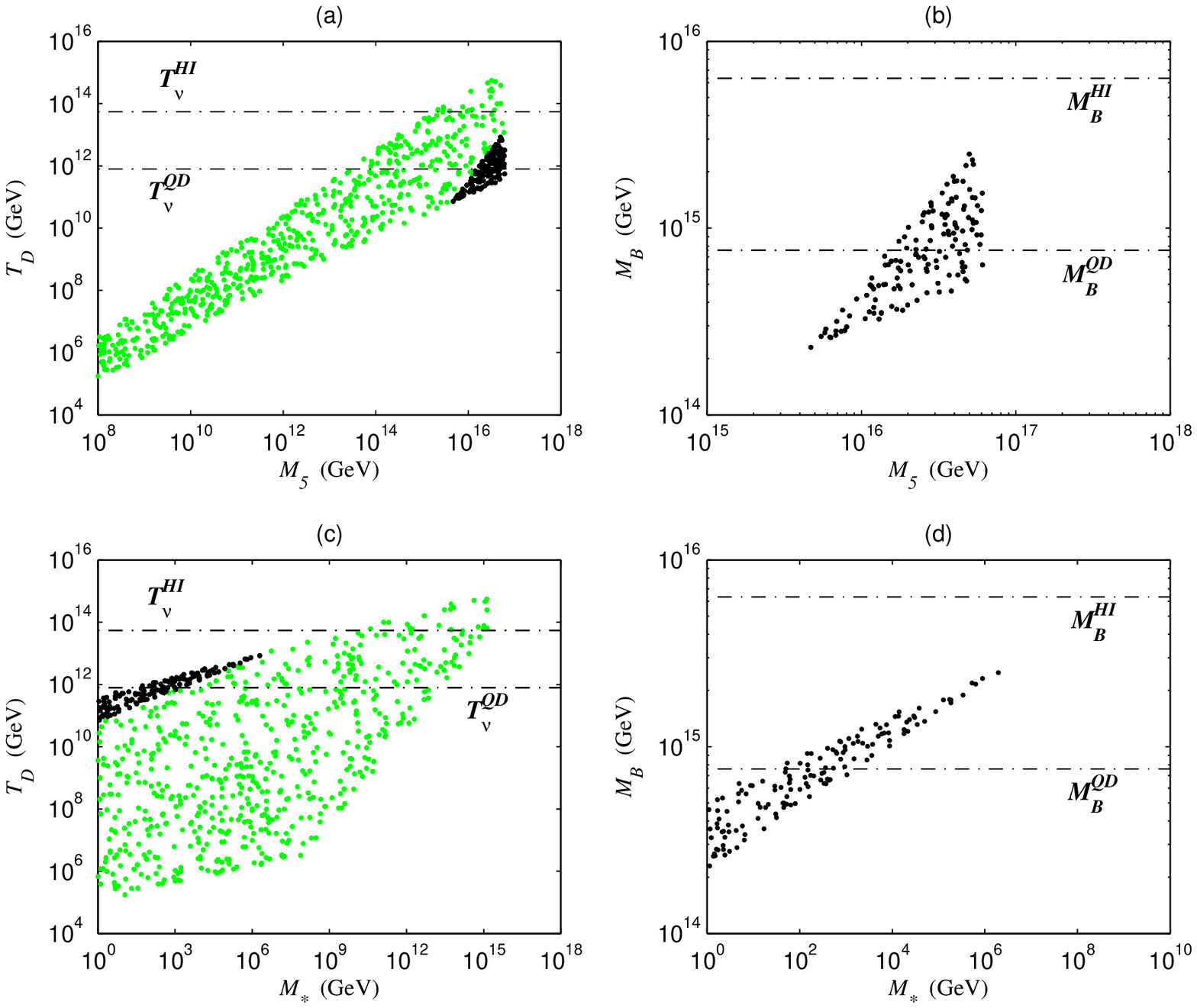}
\caption{\label{fig3} The parameter space that generates the observed
baryon-to-entropy ratio $n_B/s$ during the radiation era of Randall-Sundrum
braneworld cosmology ($\beta=0$). In this case, $M_I \approx 5 \times 10^{-2}
M_5$. For other details, see caption of Fig.~\ref{fig2}.}
\end{center}
\end{figure*}

We have also performed a random scan of the parameter space $(M_*, M_5, \beta)$
in order to find the allowed region for the gravitational baryogenesis
mechanism considered here to generate an observationally acceptable $n_B/s$
with $T_D<M_I<M_5$. The results are presented in Fig.~\ref{fig2}. A similar
analysis was done for the case when the GB terms are absent, i.e. $\beta=0$,
and braneworld cosmology is of RS type (see Fig.~\ref{fig3}). We notice that in
the RS case the gravity scale $M_5$ can take considerably lower values (cf.
Fig.~(3a)), only constrained to be larger than $10^5$~TeV, if one requires the
theory to reduce to Newtonian gravity on scales larger than 1~mm. The above
bound yields $T_D \gtrsim 10^5$~GeV.

Up to now we have not taken into account possible effects which could dilute
the baryon asymmetry generated by the mechanism described above. It is well
known that electroweak sphaleron transitions, which are unsuppressed at
temperatures above the electroweak phase transition, are a potential source of
dilution~\cite{Kuzmin:1985mm}. Sphaleron-induced baryon-asymmetry dilution
occurs when $B-L$ vanishes. In this case, the $B$ and $L$-number densities will
be typically diluted by a factor $0.02\, m_\tau^2/T_{\rm
sph}^2\,$~\cite{Bertolami:1996cq}, where $m_\tau$ is the $\tau$ lepton mass and
$T_{\rm sph}$ is the sphaleron freeze-out temperature. Assuming $T_{\rm sph}$
to be the electroweak scale, one finds that the baryon asymmetry is diluted by
a factor of about $10^{-6}$. Hence, according to Eq.~(\ref{asym}), the scale
$M_\ast$ would have to be, in this case, smaller by a factor of $10^{-3}$ to
reproduce the correct value of $n_B/s$ via the gravitational interaction of
Eq.~(\ref{current}). On the other hand, if $B-L \neq 0$, essentially no
sphaleron dilution occurs. In the latter case, the baryon asymmetry generated
will remain after the decoupling of the $(B-L)$-violating interactions. An
example of this possibility will be presented in the next section.

\section{Gravitational leptogenesis}

In the standard model of electroweak interactions, the $B-L$ symmetry is
exactly conserved. This symmetry is however violated in many of its extensions.
In general, it is possible  that the $B$-violating interactions are generated
by an operator ${\cal O}_B$ of mass dimension $D=4+n$. The rate of such
interactions is $\Gamma_B \sim T^{2 n + 1}/M_B^{2 n}$, where $M_B$ is the mass
scale associated with the operator ${\cal O}_B$. In the standard electroweak
model the lowest-dimensional operator that violates $B-L$ is the dimension five
operator
\begin{align}
{\cal L}_{\slash\!\!\!\!L} = \frac{1}{M}\, \ell\, \ell\, \phi\, \phi +
\text{H.c.}\,, \label{op}
\end{align}
where $\ell$ and $\phi$ are the left-handed lepton and Higgs doublets,
respectively; $M$ is the scale of new physics which induces $B-L$ violation.
This interaction represents a typical term that gives rise to the seesaw
mechanism and is responsible for the light neutrino masses $m_i \sim v^2/M$, $v
\simeq 174$~GeV. In the early universe the $L$-violating rate induced by the
interaction (\ref{op}) is~\cite{Fukugita:1990gb}
\begin{align} \label{gb}
\Gamma_{\slash\!\!\!\!L} = \frac{T^3}{M_B^2}\,, \quad M_B \approx \frac{10\,
v^2}{(\sum m_i^2)^{1/2}}\,.
\end{align}

The decoupling of the $(B-L)$-violating processes occurs when
$\Gamma_{\slash\!\!\!\!L}$ falls below the Hubble rate, i.e. when $ M_B \simeq
[T_D^3/H(T_D)]^{1/2}$. In the gravitational baryogenesis scenario considered
here, the decoupling temperature that produces an acceptable baryon asymmetry
is fixed by Eq.~(\ref{YB}), and it determines the required scale of $B-L$
violation. In Fig.~\ref{fig1} we have plotted $M_B$ as a function of $M_\ast$
for different values of $\beta$ and $M_5$. We notice that the requirement $T_D<
M_B < M_5$ imposes an upper bound on $M_*$. We find $M_* \lesssim 10^{16}$~GeV.

If the scale $M_B$ is associated with the neutrino mass seesaw operator, as in
Eq.~(\ref{gb}), then the value of this scale will be fixed by the light
neutrino mass spectrum. The current cosmological limit coming from the
Wilkinson Microwave Anisotropy Probe (WMAP) implies $\sum m_i \lesssim
0.69$~eV~\cite{Spergel:2003cb}. If neutrinos are quasidegenerate (QD) in mass,
the above limit requires $m_1\simeq m_2 \simeq m_3 \simeq 0.23~\mbox{eV}$. In
this case,
\begin{align}
M_B^{QD} \approx 7.6 \times 10^{14}~\mbox{GeV}\,. \label{mbqd}
\end{align}
Instead, if neutrinos masses are hierarchical (HI) with $m_1 \simeq 0 \ll m_2
\ll m_3$, then $m_2 \simeq (\Delta m_{\rm sol}^2)^{1/2}$ and $m_3 \simeq
(\Delta m_{\rm atm}^2)^{1/2}$, where the squared mass differences measured in
solar and atmospheric neutrino oscillation experiments are $\Delta m_{\rm
sol}^2 \simeq 8.1 \times 10^{-5}~\mbox{eV}^2 $ and $\Delta m_{\rm atm}^2 \simeq
2.2\times 10^{-3}~\mbox{eV}^2 $~\cite{Maltoni:2004ei}, respectively. In the
latter case,
\begin{align}
M_B^{HI} \approx 6.3 \times 10^{15}~\mbox{GeV}\,. \label{mbhi}
\end{align}
Eqs.~(\ref{mbqd}) and (\ref{mbhi}) yield a decoupling temperature in the range
\begin{align}
T_\nu^{QD} \leq T_D \leq T_\nu^{HI}\,.
\end{align}
Clearly, the specific values of $T_\nu^{QD}$ and $T_\nu^{HI}$ depend on whether
the decoupling of $B-L$ violation occurs in GB, RS or GR regime and, thus, on
the values of the Gauss-Bonnet coupling $\beta$ and the fundamental scale
$M_5$. In standard cosmology with $H(T_D) \simeq 1.66\,g_*^{1/2} T_D^2/M_4\,$,
one finds $T_\nu^{QD} \approx 7.9 \times 10^{11}~\mbox{GeV}$ and $T_\nu^{HI}
\approx 5.5 \times 10^{13}~\mbox{GeV}$. Some other examples are presented in
Fig.~\ref{fig1}. While in Fig.~(1a) and (1b), the decoupling corresponding to
$T_\nu^{QD}$ and $T_\nu^{HI}$ (horizontal dot-dashed lines) occurs in standard
cosmology, in Fig.~(1c) and (1d) such decoupling takes place in the high-energy
Gauss-Bonnet regime for the case of hierarchical neutrinos. We also remark that
for gravitational leptogenesis to be successful we must require $T_D \geq
T_\nu^{QD}$, which implies the lower bound $M_* \gtrsim 100$~GeV for $\beta
\lesssim 0.1$, as can be seen from the figure.

Let us now consider the inflation bound. Since $T_D < M_B$, the requirement
$M_B < M_I$ is more stringent in this case. We find that this bound strongly
constrains the scale of $B-L$ violation and, consequently, the mechanism of
gravitational leptogenesis. For instance, it can be seen that for the case
presented in Fig.~(1d), the above constraint implies the bound $T_D <
T_\nu^{QD}$ and, therefore, the leptogenesis mechanism cannot generate the
required baryon asymmetry. The allowed region for gravitational leptogenesis is
presented in Fig.~\ref{fig2} (black dots). The horizontal dot-dashed lines
correspond to the GR decoupling temperatures $T_\nu^{QD}$ and $T_\nu^{HI}$
(Fig.~(2c)) and the scales $M_B^{QD}$ and $M_B^{HI}$ (Fig.~(2d)) associated
with the neutrino mass seesaw operator. We conclude that
\begin{align} \label{GBbounds}
10^{15}~\mbox{GeV} & \lesssim M_5 \lesssim 10^{17}~\mbox{GeV}\,,\nonumber\\
10^{2}~\mbox{GeV} & \lesssim M_* \lesssim 10^{10}~\mbox{GeV}\,.
\end{align}

One can compare the above results with the ones that are obtained in the case
when braneworld cosmology is of RS type, i.e. when $\beta=0$. The allowed range
of values for the parameters is shown in Fig.~\ref{fig3}. We notice that a
successful gravitational leptogenesis in RS cosmology requires
\begin{align} \label{RSbounds}
10^{16}~\mbox{GeV} & \lesssim M_5 \lesssim 10^{17}~\mbox{GeV}\,,\nonumber\\
10^{2}~\mbox{GeV} & \lesssim M_* \lesssim 10^{6}~\mbox{GeV}\,.
\end{align}

\section{Conclusion}

In this work we have considered the possibility that the observed baryon
asymmetry arises via the spacetime dynamics of Gauss-Bonnet braneworld
cosmology. The framework presented here is based on the $CPT$-violating
gravitational interaction between the derivative of the Ricci scalar curvature
and the $B$ (or $B-L$) current~\cite{Davoudiasl:2004gf}. We have shown that it
is possible to generate the correct magnitude of the baryon asymmetry in
different cosmological scenarios, depending on whether the decoupling of the
$B$- or $(B-L)$-violating interactions occurs in standard cosmology or in the
high-energy Randall-Sundrum or Gauss-Bonnet braneworld regimes.

We have also studied the case when baryogenesis occurs via leptogenesis, and
the $B-L$ current is associated with the neutrino mass seesaw operator. In this
framework, the produced $n_{B-L}$ asymmetry will be converted to a baryon
asymmetry, once sphaleron transitions enter thermal equilibrium. We have seen
that for this scenario to be viable, a rather high fundamental scale of gravity
$M_5$ is required (cf. Eq.~(\ref{GBbounds})), as well as an effective
interaction scale $M_*$ above the electroweak scale but below $10^{10}$~GeV. At
this point it is worth noticing that, although in four-dimensional gravity it
is natural to expect $M_*$ of the order of the Planck mass $M_4$, this may not
be necessarily the case. For instance, $M_* \simeq (M_R M_4)^{1/2}$ could be
possible, if the right-handed neutrino Majorana mass $M_R$ softly violates
baryon number~\cite{Davoudiasl:2004gf}. Moreover, this scale can be much lower,
if the effective four-dimensional theory comes from a higher-dimensional
theory. Indeed, AdS/CFT correspondence~\cite{Maldacena:1997re} and braneworld
holography~\cite{Shiromizu:2001ve,Shiromizu:2004cb} indicate that interaction
terms such as given by Eq.~(\ref{current}) are expected in the effective action
on the brane with $M_* \simeq 1/\ell \simeq M_5^3/M_4^2$. If this is the case,
the bound $M_5 \simeq 10^{15} - 10^{17}$~GeV would then imply $M_* \simeq
10^{7} - 10^{13}$~GeV, well within the range allowed by gravitational
baryogenesis and leptogenesis in the high-energy GB regime.

\begin{acknowledgments}
We thank O. Bertolami for useful discussions and comments. The work of R.G.F.
and N.M.C.S. was supported by \emph{Funda\c{c}\~{a}o para a Ci\^{e}ncia e a Tecnologia}
(FCT, Portugal) under the grants SFRH/BPD/1549/2000 and SFRH/BD/4797/2001,
respectively.
\end{acknowledgments}

\appendix*
\section{Inflation in a Gauss-Bonnet braneworld}

In this Appendix we briefly review inflation in GB brane
cosmology~\cite{Lidsey:2003sj,Dufaux:2004qs,Tsujikawa:2004dm}. For simplicity
we assume that inflation is driven by the simple quadratic potential
\begin{align}
V(\phi)=V_0\, \phi^2 . \label{pot}
\end{align}

We are interested in slow-roll inflation which occurs in the GB or RS
high-energy regime. In this case, $V \approx \rho \gg \lambda$ and
Eq.~(\ref{chi}) implies $ V \approx m_\alpha^4 \sinh \chi$. Moreover, the bound
$V > m_\lambda^4$ together with the quantum gravity upper limit $V < M_5^4$
imply
\begin{align}
 \left(\frac{m_\lambda}{m_\alpha}\right)^4 < \sinh \chi <
 \left(\frac{M_5}{m_\alpha}\right)^4\,.
\end{align}

The slow-roll parameters $\epsilon$ and $\eta$ are given by
\begin{align}
\epsilon =\frac{16 \lambda V_0}{27 \kappa_4^2 m_\alpha^8}~ f(\chi)\,,\quad
\eta = \frac{8 \lambda V_0}{9 \kappa_4^2 m_\alpha^8}~ \frac{1}{g(\chi)}\,,
\end{align}
where
\begin{align}
f(\chi) &= g^{-2}(\chi) \tanh \chi \sinh\left(\!\frac{2\chi}{3}\!\right)\,, \\
g(\chi) &= \cosh\left(\!\frac{2\chi}{3}\!\right)-1\,.
\end{align}

The number of e-folds of inflation is given by
\begin{align}
N_\star =\frac{3\mu^2}{4\beta V_0}\int^{\chi_\star}_{\chi_e} g(\chi) \coth \chi
d\chi \equiv \left. \frac{9\mu^2}{8\beta V_0}
I(\chi)\right|_{\chi_e}^{\chi_\star},\label{efold3}
\end{align}
where
\begin{align}
I(\chi)=\,g(\chi) - \ln \left[1+ \frac{2}{3} g(\chi)\right]~,\label{func}
\end{align}
$\chi_\star$ is evaluated when cosmological scales leave the horizon and
$\chi_e$ is evaluated at the end of inflation, when ${\rm
max}\{\epsilon,\eta\}=1$.

The amplitude of scalar perturbations is
\begin{align}
A_S^2= \frac{3^{5/2}\kappa_4\mu^5}{16\pi^2V_0\lambda^{1/2}\beta^{5/2}}
\frac{g^3(\chi_\star)}{ \sinh \chi_\star} \,.\label{scalar}
\end{align}
Using the COBE normalized value $A_S \simeq 2\times 10^{-5}$ for the density
perturbations and $55 \leq N_\star \leq 65$, we can obtain the scale of
inflation $M_I=V^{1/4}(\phi_e) \approx m_\alpha \sinh^{1/4} \!\chi_e\,$. This
scale is plotted in Fig.~\ref{fig1} for given values of $\beta$ and $M_5$,
taking $N_\star=60$. A more complete analysis is presented in Fig.~(2a) and
(2b). Notice that, for consistency, one should require $m_\lambda<M_I<M_5\,$.

\end{document}